\documentclass[english,12pt]{article}
\usepackage{amsmath}
\usepackage{amsfonts}
\usepackage{latexsym}
\usepackage{graphicx}
\usepackage{multicol}
\usepackage{color}
\usepackage[usenames,dvipsnames,svgnames,table]{xcolor}
\usepackage[utf8]{inputenc}
\usepackage{amsfonts}
\begin{document}

\title{Shrinkage linear regression for symbolic interval-valued variables}

\author{Oldemar Rodr\'{\i}guez \thanks{University of Costa Rica, San Jos\'e, Costa Rica;
E-Mail: oldemar.rodriguez@ucr.ac.cr}}

\maketitle

\begin{abstract}
This paper proposes a new approach to fit a linear regression for symbolic internal-valued variables, which improves both the Center Method suggested by Billard and Diday in \cite{BillardDiday2000} and the Center and Range Method suggested by Lima-Neto, E.A. and De Carvalho, F.A.T. in \cite{Lima2008, Lima2010}. Just in the Centers Method and the Center and Range Method, the new methods proposed fit the linear regression model on the midpoints and in the half of the length of the intervals as an additional variable (ranges) assumed by the predictor variables in the training data set, but to make these fitments in the regression models, the methods Ridge Regression, Lasso, and Elastic Net proposed by Tibshirani, R. Hastie, T., and Zou H in  \cite{Tib1996, HastieZou2005} are used. The prediction of the lower and upper of the interval response (dependent) variable is carried out from their midpoints and ranges, which are estimated from the linear regression models with shrinkage generated in the midpoints and the ranges of the interval-valued predictors. Methods presented in this document are applied to three real data sets “cardiologic interval data set”, “Prostate interval data set” and “US Murder interval data set” to then compare their performance and facility of interpretation regarding the Center Method and the Center and Range Method. For this evaluation, the root-mean-squared error and the correlation coefficient are used. Besides, the reader may use all the methods presented herein and verify the results using the {\tt RSDA} package written in {\tt R} language, that can be downloaded and installed directly from {\tt CRAN}  \cite{Rod2014},
\end{abstract}

\section*{Keywords}
Interval-valued variables, Linear Regression, Elastic Net, Lasso, Ridge Regression, Symbolic Data Analysis.

\section{Introduction}
\label{sec:int}

Statistical and data mining methods have been developed mainly in the case in which variables take a single value. Nevertheless, in real life there are many situations in which the use of this type of variables may cause an important loss of information or reduction in quality and veracity of results. In the case of quantitative variables, a more complete information can be achieved by describing an ensemble of statistical units in terms of interval data, that is, when the value taken by a variable is an interval  $[a,b]$. 

An especially useful case where it is convenient to summarize large ensembles of data in such a way that the summary of data resulting is of a more manageable size, which in turn maintains the greatest amount of information it had in the original data set. In this problem, the central idea is to substitute the ensemble of all transactions carried out by a person or client (for example the owner of a credit card) for one only “transaction” that summarizes all originals in such a way that millions of transactions could be summarized in an only one that maintains the client’s habitual behavior. The above is achieved thanks to this new transaction will have in its fields not only numbers (as in the usual transactions), but will also have intervals that store, for example, the minimum and maximum purchase. In experimental evaluation section, we will provide an example that illustrates these ideas for which we will use the “US Communities and Crime Data Set”  \cite{Bache2013}.

The statistical treatment of the interval-type data has been considered in the context of Symbolic Data Analysis – SDA) introduced by E. Diday in \cite{Diday1987}, the objective of which is to extend the classic statistical methods to the study of more complex data structures that include, among others, interval-valued variables. A complete presentation on Symbolic Data Analysis can be found in the following works \cite{BockDiday2000,BillardDiday2003,BillardDiday2006}.

On the other hand, the linear models of regression were created in great measure at the time that statistics were developed without the use of computers, but even in the era of today’s computer, there are still very good reasons to continue using and investigating in the linear regression field. Regression methods are simple and frequently provide an adequate and interpretable description of how predictor variables affect the response variable. For the purposes of the quality of the predictions, the linear regression method often exceeds the more sophisticated and complex models.

A very important general problem for predictive modeling is the best subset selection of variables in order to get the best prediction. In the case of linear regression there are two reasons for this:

\begin{enumerate}
\item The first one is the exactness of prediction. The quality of prediction can sometimes be improved by shrinking or even making zero some of the coefficients of the regression.

\item The second reason is interpretation. When there is a large number of predictor variables, interpretation is much more difficult, therefore, often we would like to determine a smaller subset of variables that contains the stronger factors that impact prediction.
\end{enumerate}

There are many approaches that have been used in order to select the best subset of predictors, we will concentrate in this work in the use of Shrinkage Methods. The above because by means of the retention of a subset of predictors and disregarding the rest, usually a more interpretable model is generated and that possibly has a lesser error that if the complete model is used. This is also valid in the case of regression methods for interval-valued variables. Also, the discrete process of selecting and rejecting variables often presents a high variety in results, at the same time that the variable shrinkage process has the great advantage that since they are more continuous process, they do not suffer so much from this high variability.

We will use for this work the methods Elastic Net, Lasso and Ridge Regression to improve the methods of regression for interval-valued variables. We will use them to improve the quality of predictions of the Center Method and the Center and Range Method. The Ridge Regression shrinks coefficients or regressions by penalizing its size; coefficients are penalized when the residual sum of squares is minimized, at the same time that the Lasso regression makes also a selection and shrinkage of variables generating greater simplicity in the solution. On its part, the Elastic Network is also a method of selection of variables that can be seen as a consensus between the Lasso Method and the Ridge Regression Method, and that is particularly useful when the number $p$ of predictors is much greater than the number $n$ of observations, that is, the $p \gg n$ case.

In the frame of Symbolic Data Analysis, Rodríguez in \cite{Rod2000} introduces for the case of one single predictor, four methods to do linear regression to interval-valued variables: the simple regression with empirical correlation, linear regression based on the maximum and minimum correlation, linear regression based on the mid-points and linear regression based on top-points of the hypercubes. Subsequently Billard and Diday in \cite{BillardDiday2000} present the first approach to fit a model to general linear regression for interval-type data sets, the approach consists in fitting a model of linear regression using the midpoints of the interval-value assumed by the variables in the training table and then applying this method to predict the lower and upper values of the response interval variable. Lima Neto and De Carvalho in \cite{Lima2008} improved this approach by submitting a new method based on the generation of two linear-regression models, the first one is fitted with the medium points of intervals and the second one is fitted with ranges of the intervals. Which permits the reconstruction of the limits of the values of the response interval variable based on the sums and subtractions of the values generated by these two regression models, improving in many cases in an important manner the quality of predictions. Finally, Lima Neto and De Carvalho in \cite{Lima2010}  suggest the constrained linear regression model for internal-valued data. This new approach to fit a linear regression model makes an important improvement to the Center and Range Method using restrictions expressed as inequalities in the model’s parameters to mathematically guarantee that the inferior limit be always less than the superior limit in the interval of the response variable that is being subject of prediction.

In this work, we concentrate in improving the Center Method and the Center and Range Method. To this effect, in section \ref{sec:mlc} we will make a summarized presentation of the linear regression models to quantitative variables single valued. In section \ref{sec:mli}, we present a summary of the Center Method and the Center and Range Method. In section \ref{sec:mlci} we suggest six new models of linear regression for interval-valued variables that improve the Center Method and the Center and Range Method. Finally, in section \ref{sec:ee} we present an experimental evaluation with three real data sets that evidence important improvements in the results. 

\section{Shrinkage Linear Regresion Methods}
\label{sec:mlc}

In this section we present a summary of the shrinkage linear regression models, a complete presentation can be found in \cite{HastieTib2008} and \cite{HastieZou2005}. Linear regression models have been for decades one of the most important predictive methods in statistics; and in fact, it continues being today one of the most important tools in Statistics and Mining Data. As it is well-known, the idea is, given an input vector, $x^t = (x_1, x_2,\ldots, x_p)$, we want to predict the response variable $y$ through the following linear regression model:

\begin{equation} \label{eq:r1}
\hat{y} = \hat{\beta_0} + \sum_{j=1}^{p} x_j \hat{\beta_j},
\end{equation}

\noindent the term $\beta_0$ is denominated the intercept, also known as the bias in machine-learning. If a constant 1 is included to vector  $x$ and $\beta_0$ in the coefficients’ vector $\beta$ the linear model can be written in vectorial form as a product as follows:

\begin{equation} \label{eq:r2}
\hat{y} = x^t \hat{\beta}, 
\end{equation}

\noindent where $x^t$ denotes a transpose vector, being $x$ a column vector. To fit the linear model in the training data, the most popular estimation method is least squares. In this approach, we pick the coefficients $\beta$ to minimize the residual sum of squares:

\begin{equation} \label{eq:r3}
\text{RSS}(\beta) =   \sum_{i=1}^{n} \left(y_i-x_i^t \beta \right)^2.
\end{equation}

$\text{RSS}(\beta)$ is a quadric function; therefore, its minimum always exists.  It can be written as:

\begin{equation} \label{eq:r4}
\text{RSS}(\beta) =  (y-X \beta)^t (y-X \beta),
\end{equation}

\noindent  where $X$ is matrix $n \times p$ wherein each row is a vector in the training data set; and $y$ is an $n$ size vector (the output vector in the training data set). It is well-known that if $X^t X$ is nonsingular matrix the solution is given by: 

\begin{equation} \label{eq:r5}
\hat{\beta} =  (X^t X)^{-1}X^t y,
\end{equation}

\noindent  the approximate value by this model for the $i-$teenth component $x_i$  can be estimated as $\hat{y}_i=x_i^t \hat{\beta}$ and the fitted values for a new case $x^t=(1,x_1,\ldots,x_p)$ is given by $\hat{y}=x^t \hat{\beta}$.

In the case of ridge regression, the coefficients of the regression model are shrunk by imposing a penalization to its size. The coefficients in this case are obtained minimizing the penalized residual sum of squares, that is, instead of minimizing (\ref{eq:r3}), the following is minimized: 

\begin{equation} \label{eq:r6}
\hat{\beta}^{\text{ridge}} =  \underset{\beta}{\operatorname{argmin}}  \left\{ \sum_{i=1}^{n} \Big((y_i - \beta_0 - \sum_{j=1}^{p} x_{ij} \beta_j \Big)^2 + \lambda \sum_{j=1}^{p} \beta_j^2 \right\},
\end{equation}

\noindent where $\lambda \ge 0 $ is the complexity parameter that controls the amount of shrinkage, thus the larger the value of $\lambda $, the greater will the amount of the shrinkage of the regression coefficients. It is known that the expression (\ref{eq:r6}) can be written equivalently as:

\begin{eqnarray} \label{eq:r7}
\hat{\beta}^{\text{ridge}} &=&  \underset{\beta}{\operatorname{argmin}}   \sum_{i=1}^{n} \left(y_i - \beta_0 - \sum_{j=1}^{p} x_{ij} \beta_j \right)^2, \\
& & \text{subject to }   \sum_{j=1}^{p} \beta_j^2  \le t, \nonumber
\end{eqnarray}

\noindent which makes explicit the restriction of the size of the parameters. There is one to one correspondence between the parameters $\lambda$ in (\ref{eq:r6}) and $t$ in (\ref{eq:r7}). Equation (\ref{eq:r6}) can be written in matrix form as follows:

\begin{equation} \label{eq:r8}
\text{RSS}(\beta) =  (y-X \beta)^t (y-X \beta) + \lambda \beta^t \beta,
\end{equation}

\noindent with this expression, it can easily be proven that:

\begin{equation} \label{eq:r9}
\hat{\beta}^{\text{ridge}} = ( X^t X + \lambda I )^{-1} X^t  y,
\end{equation}

\noindent where $I$ is the identity matrix $p \times p$. Observe that this solution of the regression with shrinkage is again a linear function of $y$, the solution also adds a positive constant to the diagonal of $X^t X$ before the estimate of its inverse, which again makes that this be a nonsingular problem. In a similar way, the approximate value by this model for the $i-$teenth component $x_i$  can be estimated as $\hat{y}_i=x_i^t \hat{\beta}^{\text{ridge}}$ and the fitted values for a new case $x^t=(1,x_1,\ldots,x_p)$ is given by $\hat{y}=x^t \hat{\beta}^{\text{ridge}}$.

The lasso method was proposed by Tibshirani, R. in \cite{Tib1996}. This method is also a shrinking method similar to the ridge regression method, the difference is that in ridge regression a penalization type $L_2$ is used with the form $ \sum_{j=1}^{p} \beta_j^2$, while the lasso regression uses penalization type $L_1$ with the form $ \sum_{j=1}^{p} | \beta_j |$. The lasso regression problem can then be written in its Lagrangian form as:

\begin{equation} \label{eq:r10}
\hat{\beta}^{\text{lasso}} =  \underset{\beta}{\operatorname{argmin}}  \left\{ \sum_{i=1}^{n} \Big(y_i - \beta_0 - \sum_{j=1}^{p} x_{ij} \beta_j \Big)^2 + \lambda \sum_{j=1}^{p} | \beta_j | \right\},  
\end{equation}

\noindent or its equivalent as:

\begin{eqnarray} \label{eq:r11}
\hat{\beta}^{\text{lasso}} &=&  \underset{\beta}{\operatorname{argmin}}   \sum_{i=1}^{n} \left(y_i - \beta_0 - \sum_{j=1}^{p} x_{ij} \beta_j \right)^2, \\
& & \text{subject to }   \sum_{j=1}^{p} | \beta_j |  \le t. \nonumber
\end{eqnarray}

In this case, this type of penalizing the use of the absolute value makes that the solution not be a linear function of $y$ and therefore the solution doesn't have a closed form as in the case of the ridge regression. Estimate the solution at (\ref{eq:r11}) is a quadratic programming problem to which effect very efficient algorithms have been found that permit the estimation of the complete path of solution when $\lambda$ varies. For example, in Osborne et al. \cite{Osborne2000} had proposed an efficient piecewise-linear algorithm known as a homotopy algorithm.

Due to the nature of the restrictions in (\ref{eq:r11}) when $t$ becomes sufficiently small (or equivalently, when $\lambda$  increase in (\ref{eq:r10})), it results that some of the coefficients of regression solution become exactly zero, this makes the lasso method a continuous selection method of variables. The lasso regression method is a method significantly different from the ridge regression since it converts it into a method that automatically makes variables selection generating sparse models and solutions easier to interpret. Also, as we will see in section \ref{sec:ee} in terms of prediction error, the lasso regression method is in many cases better than the ridge regression method.

Analogous to the above methods, the approximate value by this model for the $i-$teenth component $x_i$  can be estimated as $\hat{y}_i=x_i^t \hat{\beta}^{\text{lasso}}$ and the fitted values for a new case $x^t=(1,x_1,\ldots,x_p)$ is given by $\hat{y}=x^t \hat{\beta}^{\text{lasso}}$.

The elastic net penalty method was proposed by Zou and Hastie in the paper “Regularization and variable selection via the elastic net” \cite{HastieZou2005}. This method proposes a consensus between the penalizations of the ridge regression and the lasso method, thus, the penalization factor has the form:

\begin{eqnarray} \label{eq:r12}
\alpha \sum_{j=1}^{p} | \beta_j | + (1-\alpha) \sum_{j=1}^{p} \beta_j^2.
\end{eqnarray}

The elastic-net selects variables just like the lasso method and at the same time shrink the coefficients of the predictors that are highly correlated. Observe also that the second term of this consensus furthers that highly correlated variables be averaged while the first term furthers sparse solution in the coefficients of these averaged variables, since it tends to cause that some of the coefficients of the regression model be zero. Thus, the elastic net penalty method is expressed as follows:

\begin{eqnarray} \label{eq:r13}
\hat{\beta}^{\text{net}} = \underset{\beta}{\operatorname{argmin}} \left\{ \sum_{i=1}^{n} \Big(y_i-\beta_0- \sum_{j=1}^{p} x_{ij} \beta_j \Big)^2 + \lambda \left( \alpha \sum_{j=1}^{p} | \beta_j | + (1-\alpha) \sum_{j=1}^{p} \beta_j^2 \right) \right\} ,
\end{eqnarray}

\noindent so, if $\alpha=0$, the elastic net becomes ridge regression, while if $\alpha=1$, the elastic net becomes lasso regression. Just like the lasso method, in the elastic net method, the solution is not a linear function of $y$, and therefore the solution does not have a closed form, in virtue of which to compute the solution to this problem a quadratic programming problem should be solved. 
The approximate value by this model for the $i-$teenth component $x_i$  can be estimated as $\hat{y}_i=x_i^t \hat{\beta}^{\text{net}}$ and the fitted values for a new case $x^t=(1,x_1,\ldots,x_p)$ is given by $\hat{y}=x^t \hat{\beta}^{\text{net}}$.

In section \ref{sec:mlci} we will generalize the application of all those shrinkage methods to the case then both predictors and response are interval-valued variables.

\section{Linear regression models for symbolic interval-valued variables}
\label{sec:mli}

In this section we will present, summarized, the center method proposed by Billard and Diday, a complete presentation can be found in \cite{BillardDiday2000}. We will also present, summarized, the center and range method proposed by Lima Neto and De Carvalho, a complete presentation of this method can  be  \cite{Lima2008} and \cite{Lima2010}.

\subsection{Center method}

In the center method, the parameters $\beta$ are estimated based on the intervals midpoints. In this method there are predictors $X_1, \ldots ,X_p$  and a response to be predicted $Y$, all these are interval-valued. Thus $X$ is a $n \times p$ matrix where each line is a vector of components of the training data set $x_i = (x_{i1}, \ldots,x_{ip})$ with  $x_{ij}=[a_{ij},b_{ij}]$ and each component of the $Y$ variable is also an interval $y_i = [y_{Li},y_{Ui}]$.

We denote by $X^c$ the matrix with the intervals midpoints of the matrix $X$, that is, $x^c_{ij}=(a_{ij}+b_{ij})/2$ and by $y^c_i=(y_{Li}+y_{Ui})/2$ the midpoints of  $Y$, then idea of the center method is to fit a linear regression model over $X^c=((x^c_1)^t,\ldots,(x^c_n)^t))^t$ with $(x^c_i)^t=(1,x^c_{i1}, \ldots,x^c_{ip})$ for $i=1,\ldots,n$ and $y^c=(y^c_1,\ldots,y^c_n)^t$. If $(X^c)^t X^c$ is nonsingular from (\ref{eq:r5}) we know that the unique solution for $\beta$ is given by:

\begin{equation} \label{eq:ri1}
\hat{\beta} =  ((X^c)^t X^c)^{-1}(X^c)^t y^c,
\end{equation}

\noindent The value of the prediction for $y=[y_L,y_U]$ for a new case $x=(x_1,\ldots,x_p)$ with $x_j=[a_j,b_j]$ is estimated as follows:

\begin{equation} \label{eq:ri2}
\hat{y}_L=(x_L)^t \hat{\beta} \;\;\; \text{ y } \;\;\; \hat{y}_U=(x_U)^t \hat{\beta},
\end{equation}

\noindent donde $(x_L)^t=(1,a_1,\ldots,a_p) $  y $(x_U)^t=(1,b_1,\ldots,b_p)$.

\subsection{Center and range method}

In center and range method, Lima Neto and De Carvalho propose a new approach to fit the linear regression model for interval-valued variables using the information contained in the midpoints and in the interval ranges, in order to improve the quality of prediction of the centers method. The idea is to fit two regression models, the first one with the midpoint of the interval and the second with the ranges of those same intervals. Just like the center method, there are $X_1, \ldots ,X_p$ predictors and a response $Y$, all these variables are interval-valued. Thus, $X$ is a $n \times p$  matrix where each row is a vector of component of the training data set $x_i = (x_{i1}, \ldots,x_{ip}) $ with $x_{ij}=[a_{ij},b_{ij}]$ and each component of the variable $Y$ is also an interval $y_i = [y_{Li},y_{Ui}]$.

To fit the first regression model, we proceed in the same way as in the center method, that is, if we denote by $X^c$ the midpoints matrix, that is, $x^c_{ij}=(a_{ij}+b_{ij})/2$ and by $y^c_i=(y_{Li}+y_{Ui})/2$ the midpoints of $Y$, the center and range method fit a first linear regression model over $X^c=((x^c_1)^t,\ldots,(x^c_n)^t))^t$  with $(x^c_i)^t=(1,x^c_{i1},\ldots,(x^c_n)^t))^t$ for $i=1,\ldots,n$ y $y^c=(y^c_1,\ldots,y^c_n)^t$. Also in this case if $(X^c)^t X^c$ is nonsingular, then we know that unique solution for $\beta^c$ is given by:

\begin{equation} \label{eq:ri3}
\hat{\beta}^c =  ((X^c)^t X^c)^{-1}(X^c)^t y^c.
\end{equation}

To fit the second regression model the half of the value of the range of each interval is used. For this, we denote by $X^r$ the matrix that contains in each component half of the of interval ranges of the matrix $X$, i.e.  $x^r_{ij}=(b_{ij}-a_{ij})/2$  and by $y^r_i=(y_{Ui}-y_{Li})/2$ the half of the interval-valued variable $Y$, the center and range method fits a second linear regression model over $X^r=((x^r_1)^t,\ldots,(x^r_n)^t))^t$  with $(x^r_i)^t=(1,x^r_{i1}, \ldots,x^r_{ip})$ for $i=1,\ldots,n$ and $y^r=(y^r_1,\ldots,y^r_n)^t$ Also, in this case if $(X^r)^t X^r$ is nonsingular from equation (\ref{eq:r5}) we know that the solution for $\beta^r$ is given by:

\begin{equation} \label{eq:ri4}
\hat{\beta}^r =  ((X^r)^t X^r)^{-1}(X^r)^t y^r,
\end{equation}

\noindent thus each case in the training data set is represented by two vectors $w_i=(x^c_i,y^c_i)$ and  $r_i=(x^r_i,y^r_i)$ for $i=1,\ldots,n$. Then the prediction value for $y=[y_L,y_U]$ for a new case $x=(x_1,\ldots,x_p)$ with $x_j=[a_j,b_j]$  is estimated as follows:

\begin{equation} \label{eq:ri5}
\hat{y}_L= \hat{y}^c - \hat{y}^r \;\;\; \text{ and } \;\;\; \hat{y}_U= \hat{y}^c + \hat{y}^r ,
\end{equation}

\noindent with 

\begin{equation} \label{eq:ri6}
\hat{y}^c=(x^c)^t \hat{\beta}^c \;\;\; \text{ and } \;\;\; \hat{y}^r=(x^r)^t \hat{\beta}^r,
\end{equation}

\noindent where  $(x^c)^t=(1,x^c_1,\ldots,x^c_p) $ and $(x^r)^t=(1,x^r_1,\ldots,x^r_p)$.

This model has the problem, explained and improved by Lima Neto and De Carvalho in \cite{Lima2010} that it cannot be mathematically guaranteed that $\hat{y}_{Li} \le \hat{y}_{Ui}$ for all $i=1,\ldots n$.

\section{Shrinkage linear regression for symbolic interval-valued variables}
\label{sec:mlci}

In this section we propose six new linear regression models which, based on the shrinking methods, improve the quality of prediction of the center method and center and range method. Just like in the center method, for all methods proposed in this section, we have $p$ predictors and a response $Y$, all these variables are internal-valued. Thus, $X$ is a $n \times p$ matrix where each row is a vector of components of the training data set  $x_i = (x_{i1}, \ldots,x_{ip})  $ with $x_{ij}=[a_{ij},b_{ij}]$ and each component of the response $Y$ is also an interval $y_i = [y_{Li},y_{Ui}]$. We also denote by $X^c$ the midpoints matrix, that is, $x^c_{ij}=(a_{ij}+b_{ij})/2$ and by $y^c_i=(y_{Li}+y_{Ui})/2$ the midpoints of $Y$.

\subsection{Shrinkage center methods}

The basic idea of the Ridge Center Method is to estimate the $\beta$ parameter based on the of interval midpoints, but using for this ridge regression, then parameters $\beta$ are found by means of the resolution of the following optimization problem:

\begin{equation} \label{eq:ris1}
\hat{\beta}^{\text{c-ridge}} =  \underset{\beta}{\operatorname{argmin}}  \left\{ \sum_{i=1}^{n} \Big((y^c_i - \beta_0 - \sum_{j=1}^{p} x^c_{ij} \beta_j \Big)^2 + \lambda \sum_{j=1}^{p} \beta_j^2 \right\}. 
\end{equation}

The idea of the ridge center method is to fit a ridge regression model over $X^c=((x^c_1)^t,\ldots,(x^c_n)^t))^t$ with $(x^c_i)^t=(1,x^c_{i1}, \ldots,x^c_{ip})$ for $i=1,\ldots,n$ y $y^c=(y^c_1,\ldots,y^c_n)^t$.  This case, if the matrix $((X^c)^t X^c + \lambda I)$ is nonsingular, we know that the solution for $\beta$ is given by:

\begin{equation} \label{eq:ris2}
\hat{\beta}^{\text{c-ridge}} = ( (X^c)^t X^c + \lambda I )^{-1} (X^c)^t y^c,
\end{equation}

\noindent where $I$ is  the identity matrix $p \times p$.  The value of the prediction for $y=[y_L,y_U]$ to a new case $x=(x_1,\ldots,x_p)$ with $x_j=[a_j,b_j]$ is estimated as follows:

\begin{equation} \label{eq:ris3}
\hat{y}_L=(x_L)^t \hat{\beta}^{\text{c-ridge}} \;\;\; \text{ y } \;\;\; \hat{y}_U=(x_U)^t \hat{\beta}^{\text{c-ridge}},
\end{equation}

\noindent where $(x_L)^t=(1,a_1,\ldots,a_p) $ and $(x_U)^t=(1,b_1,\ldots,b_p)$. 

For the Lasso Center Method the idea is similar, in this case, parameters  $\beta$ are found by means of the resolution of the following problem of optimization:

\begin{equation} \label{eq:ris4}
\hat{\beta}^{\text{c-lasso}} =  \underset{\beta}{\operatorname{argmin}}  \left\{ \sum_{i=1}^{n} \Big((y^c_i - \beta_0 - \sum_{j=1}^{p} x^c_{ij} \beta_j \Big)^2 + \lambda \sum_{j=1}^{p} | \beta_j | \right\},
\end{equation}

\noindent as we know, in this type of penalization the solution is not a linear function of $y^c_i$, and therefore the solution doesn't have a closed form, therefore to estimate the solution to this problem in each case a quadratic programming problem is solved. The value of the prediction for $y=[y_L,y_U]$ for a new case $x=(x_1,\ldots,x_p)$ with $x_j=[a_j,b_j]$ is estimated as follows:

\begin{equation} \label{eq:ris5}
\hat{y}_L=(x_L)^t \hat{\beta}^{\text{c-lasso}} \;\;\; \text{ y } \;\;\; \hat{y}_U=(x_U)^t \hat{\beta}^{\text{c-lasso}},
\end{equation}

\noindent where $(x_L)^t=(1,a_1,\ldots,a_p) $  y $(x_U)^t=(1,b_1,\ldots,b_p)$. 

For the Elastic Net Center Method the  $\beta$ parameters are found by means of the resolution of the following optimization problem:

\begin{equation} \label{eq:ris6}
\hat{\beta}^{\text{c-net}} = \underset{\beta}{\operatorname{argmin}} \left\{ \sum_{i=1}^{n} \Big(y^c_i-\beta_0- \sum_{j=1}^{p} x^c_{ij} \beta_j \Big)^2 + \lambda \left( \alpha \sum_{j=1}^{p} | \beta_j | + (1-\alpha) \sum_{j=1}^{p} \beta_j^2 \right) \right\} ,
\end{equation}

\noindent where if $\alpha=0$, the elastic net center method becomes ridge center method regression, while if $\alpha=1$, the elastic net becomes lasso center regression. Just like in the lasso center method, in the elastic net center the solution is not a lineal function of  $y$ and therefore, the solution to each problem in each case should also solve a quadratic programming problem. Besides, the value of prediction for $y=[y_L,y_U]$ for a new case $x=(x_1,\ldots,x_p)$ with $x_j=[a_j,b_j]$ is estimated as follows:

\begin{equation} \label{eq:ris7}
\hat{y}_L=(x_L)^t \hat{\beta}^{\text{c-net}} \;\;\; \text{ y } \;\;\; \hat{y}_U=(x_U)^t \hat{\beta}^{\text{c-net}},
\end{equation}

\noindent where $(x_L)^t=(1,a_1,\ldots,a_p) $  y $(x_U)^t=(1,b_1,\ldots,b_p)$. 

As we will see in section \ref{sec:ee}, for all method proposed in this section, in the practice, the optimum value of $\lambda$  will be estimated using cross-validation. 

\subsection{Shrinkage center and range methods}

The application of the shrinkage methods to the center and range method results more complicated since in this case two different models of regression are fitted, one for the midpoints and one for the ranges. The problem arises when there is a selection of variables, as in the case of the lasso method, because if a set of variables is chosen to the model that is fitted with the midpoints and another set of variables is selected for the model that fits the ranges the interpretation of the model as a whole would be very difficult.

The idea in Ridge Center and Range Method, just like the center and range method, is to fit two models of regression, the first one with the interval midpoints and the second one with the ranges of those same intervals. To fit the first regression model, the same procedure as in the center method is followed, that is, with the same notation of the above sections, parameters $\beta$ are found by resolving the following optimization problem:

\begin{equation} \label{eq:ris8}
\hat{\beta}^{\text{c-ridge}} =  \underset{\beta}{\operatorname{argmin}}  \left\{ \sum_{i=1}^{n} \Big((y^c_i - \beta_0 - \sum_{j=1}^{p} x^c_{ij} \beta_j \Big)^2 + \lambda \sum_{j=1}^{p} \beta_j^2 \right\}, 
\end{equation}

\noindent as we know if the matrix $((X^c)^t X^c + \lambda I)$ is nonsingular  the solution for $\beta$ is given by:

\begin{equation} \label{eq:ris2}
\hat{\beta}^{\text{c-ridge}} = ( (X^c)^t X^c + \lambda I )^{-1} (X^c)^t y^c.
\end{equation}

The second regression model fits over the interval ranges, thus the parameters $\beta$ for the regression model of the ranges is estimated by means of resolution of the following optimization problem:

\begin{equation} \label{eq:ris9}
\hat{\beta}^{\text{r-ridge}} =  \underset{\beta}{\operatorname{argmin}}  \left\{ \sum_{i=1}^{n} \Big((y^r_i - \beta_0 - \sum_{j=1}^{p} x^r_{ij} \beta_j \Big)^2 + \lambda \sum_{j=1}^{p} \beta_j^2 \right\},
\end{equation}

\noindent also in this case if $(X^r)^t X^r + \lambda I )$ is nonsingular the only solution for $\beta^\text{r-ridge} $ is given by:

\begin{equation} \label{eq:ris10}
\hat{\beta}^\text{r-ridge} =   ( (X^r)^t X^r + \lambda I )^{-1} (X^r)^t y^r,
\end{equation}

\noindent Thus, analogously to the and center and range method, each case in the training data set is represented by two vectors
 $w_i=(x^c_i,y^c_i)$ and $r_i=(x^r_i,y^r_i)$ for $i=1,\ldots,n$.  Then the prediction to $y=[y_L,y_U]$ for new case $x=(x_1,\ldots,x_p)$ with $x_j=[a_j,b_j]$ is estimated as follows:

\begin{equation} \label{eq:ris11}
\hat{y}_L= \hat{y}^{\text{c-ridge}} - \hat{y}^{\text{r-ridge}} \;\;\; \text{ y } \;\;\; \hat{y}_U= \hat{y}^{\text{c-ridge}} + \hat{y}^{\text{r-ridge}} ,
\end{equation}

\noindent with 

\begin{equation} \label{eq:ris12}
\hat{y}^{\text{c-ridge}}=(x^c)^t \hat{\beta}^{\text{c-ridge}} \;\;\; \text{ y } \;\;\; \hat{y}^{\text{r-ridge}}=(x^r)^t \hat{\beta}^{\text{r-ridge}},
\end{equation}

\noindent where  $(x^c)^t=(1,x^c_1,\ldots,x^c_p) $  y $(x^r)^t=(1,x^r_1,\ldots,x^r_p)$.

To generalize the center and range method for interval-valued variables using the lasso method, also two models of regression have to be fitted. The first one for the midpoints and the second one for the ranges of intervals. However, this second model is fitted only using the variables that the first lasso model selected over the midpoints, that is, it is done only over those variables for which the $\beta_i^{\text{c-lasso}}$  is not zero. The above because if two models are generated independently, each one would make its own selection of variables; and actually, a prediction could be estimated with these two models, but its interpretation would be very difficult since the two models would be using a different set of predictors.

To fit the first lasso regression model of over the midpoints, parameters $\beta$ are found by means of the following optimization problem: 

\begin{equation} \label{eq:ris13}
\hat{\beta}^{\text{c-lasso}} =  \underset{\beta}{\operatorname{argmin}}  \left\{ \sum_{i=1}^{n} \Big((y^c_i - \beta_0 - \sum_{j=1}^{p} x^c_{ij} \beta_j \Big)^2 + \lambda \sum_{j=1}^{p} | \beta_j | \right\},
\end{equation}

The fitment of the second lasso regression model over the of the interval ranges is made restricted to the predictors selected by the first regression model. This could even impair slightly the quality of the prediction; but it is done this way in order to generate a model of easier interpretation. Thus, parameters $\beta$ are found through the resolution of the following optimization problem:

\begin{equation} \label{eq:ris14}
\hat{\beta}^{\text{r-lasso}} =  \underset{\beta}{\operatorname{argmin}}  \left\{ \sum_{i=1}^{n} \Big((y^r_i - \beta_0 -  \hspace{-3mm} \sum_{\substack{j=1 \\ \hat{\beta}_j^{\text{c-lasso}} \neq 0 }}^{p} \hspace{-3mm} x^r_{ij} \beta_j \Big)^2 + \lambda \hspace{-3mm}  \sum_{\substack{j=1 \\ \hat{\beta}_j^{\text{c-lasso}} \neq 0} }^{p} \hspace{-3mm}  | \beta_j | \right\},
\end{equation}

\noindent that is, to fit this second lasso regression model only the variables selected by the first model are used, therefore, $\hat{\beta}_j^{\text{r-lasso}}=0$ if $\hat{\beta}_j^{\text{c-lasso}}=0$ for $j=1,\ldots,p$. The value of the prediction for $y=[y_L,y_U]$ for a new case  $x=(x_1,\ldots,x_p)$ with $x_j=[a_j,b_j]$ is estimated as follows:

\begin{equation} \label{eq:ris15}
\hat{y}_L= \hat{y}^{\text{c-lasso}} - \hat{y}^{\text{r-lasso}} \;\;\; \text{ and } \;\;\; \hat{y}_U= \hat{y}^{\text{c-lasso}} + \hat{y}^{\text{r-lasso}} ,
\end{equation}

\noindent with 

\begin{equation} \label{eq:ris16}
\hat{y}^{\text{c-lasso}}=(x^c)^t \hat{\beta}^{\text{c-lasso}} \;\;\; \text{ and } \;\;\; \hat{y}^{\text{r-lasso}}=(x^r)^t \hat{\beta}^{\text{r-lasso}},
\end{equation}

\noindent where  $(x^c)^t=(1,x^c_1,\ldots,x^c_p) $  y $(x^r)^t=(1,x^r_1,\ldots,x^r_p)$.

To generalize, elastic net center and range method, just like with the lasso center and range method, two regression models are fitted, the first one for the interval midpoints and the second for the interval ranges. Also, this second model is fitted using the variables that the first model elastic net center and range method selected on the midpoints. This because the elastic net method depending on the values chosen for parameters $\alpha$ y $\lambda$ could also make variables selection.

To fit the first elastic net model over midpoints, parameters $\beta$ are found through the resolution of the following optimization problem:

\begin{equation} \label{eq:ris17}
\hat{\beta}^{\text{c-net}} = \underset{\beta}{\operatorname{argmin}} \left\{ \sum_{i=1}^{n} \Big(y^c_i-\beta_0- \sum_{j=1}^{p} x^c_{ij} \beta_j \Big)^2 + \lambda \left( \alpha \sum_{j=1}^{p} | \beta_j | + (1-\alpha) \sum_{j=1}^{p} \beta_j^2 \right) \right\} ,
\end{equation}

The fitment of the second elastic net method model over the ranges of the intervals, just like in the lasso center and range method, it is done restricted to the variables that selected the first regression model. Thus, parameters $\beta$ are found by resolving the following optimization problem:

\begin{equation} \label{eq:ris18}
\hat{\beta}^{\text{r-net}} = \underset{\beta}{\operatorname{argmin}} \left\{ \sum_{i=1}^{n} \Big(y^r_i-\beta_0- \hspace{-3mm} \sum_{\substack{j=1 \\ \hat{\beta}_j^{\text{c-lasso}} \neq 0 }}^{p} \hspace{-3mm} x^r_{ij} \beta_j \Big)^2 + \lambda \left( \alpha \hspace{-3mm} \sum_{\substack{j=1 \\ \hat{\beta}_j^{\text{c-lasso}} \neq 0 }}^{p} \hspace{-3mm} | \beta_j | + (1-\alpha) \hspace{-3mm} \sum_{\substack{j=1 \\ \hat{\beta}_j^{\text{c-lasso}} \neq 0 }}^{p} \hspace{-3mm} \beta_j^2 \right) \right\} ,
\end{equation}

\noindent so $\hat{\beta}_j^{\text{r-net}}=0$  if $\hat{\beta}_j^{\text{c-net}}=0$ para $j=1,\ldots,p$. The value of the prediction for $y=[y_L,y_U]$ for a new case $x=(x_1,\ldots,x_p)$ with $x_j=[a_j,b_j]$  is estimated as follows:

\begin{equation} \label{eq:ris19}
\hat{y}_L= \hat{y}^{\text{c-net}} - \hat{y}^{\text{r-net}} \;\;\; \text{ y } \;\;\; \hat{y}_U= \hat{y}^{\text{c-net}} + \hat{y}^{\text{r-net}} ,
\end{equation}

\noindent with 

\begin{equation} \label{eq:ris20}
\hat{y}^{\text{c-net}}=(x^c)^t \hat{\beta}^{\text{c-net}} \;\;\; \text{ y } \;\;\; \hat{y}^{\text{r-net}}=(x^r)^t \hat{\beta}^{\text{r-net}},
\end{equation}

\noindent where  $(x^c)^t=(1,x^c_1,\ldots,x^c_p) $  y $(x^r)^t=(1,x^r_1,\ldots,x^r_p)$.

All methods proposed in this section have the problem of the method of the centers and ranges explained and improved by Lima Neto and De Carvalho in \cite{Lima2010} that it cannot be guaranteed that mathematically $\hat{y}_{Li} \le \hat{y}_{Ui}$  for every $i=1,\ldots n$. In future works, we will be resolving this problem.

\section{Experimental evaluation}
\label{sec:ee}

As done by Lima Neto and De Carvalho in \cite{Lima2010}  the evaluation of the results of these linear regression models for interval-valued variables is carried out using the following indexes: the lower boundary root-mean-square-error (RMSE$_L$), the upper boundary root-mean-square error (RMSE$_U$), the square of the lower boundary correlation coefficient ($r^2_L$) and the square of the upper boundary correlation coefficient ($r^2_U$) defined as follows:

\begin{equation} \label{eq:ee1}
\text{RMSE}_L=\sqrt{\frac{\displaystyle{\sum_{i=1}^{n} (y_{Li}-\hat{y}_{Li})^2}}{n}} \;\;\; \text{and} \;\;\; 
\text{RMSE}_U=\sqrt{\frac{\displaystyle{\sum_{i=1}^{n} (y_{Ui}-\hat{y}_{Ui})^2}}{n}} ,
\end{equation}

\begin{equation} \label{eq:ee2}
r^2_L= \left(\frac{\text{Cov}(y_L,\hat{y}_L)}{S_{y_L} S_{\hat{y}_L}} \right)^2 \;\;\; \text{and} \;\;\; 
r^2_U= \left(\frac{\text{Cov}(y_U,\hat{y}_U)}{S_{y_U} S_{\hat{y}_U}} \right)^2 ,
\end{equation}

\noindent where  $y_i=[y_{Li},y_{Ui}]$ and its corresponding prediction is $\hat{y}_i=[\hat{y}_{Li},\hat{y}_{Ui}]$ for $i=1,\ldots,n$, $y_L=(y_{L1},\ldots,y_{Ln})^t$, $\hat{y}_L=(\hat{y}_{L1},\ldots,\hat{y}_{Ln})^t$,  $y_U=(y_{U1},\ldots,y_{Un})^t$, $\hat{y}_U=(\hat{y}_{U1},\ldots,\hat{y}_{Un})^t$; as is usual $\text{Cov}(\Psi,\Phi)$ denotes the covariance among variables $\Psi$ and $\Phi$; and $S_{\Psi}$ denotes the standard deviation of variable $\Psi$.

All examples presented in this section were processed using the package {\tt RSDA} ({\tt R} to Symbolic Data Analysis) constructed by the author of this work for applications of the Symbolic Data Analysis, can be consulted in \cite{Rod2014}. The data sets used are also contained in package {\tt RSDA}.

\subsection{Cardiological interval data set}

This data set contains registries for three inter-valued predictors ``pulse rate'' $Y$, systolic blood pressure $X_1$ and diastolic blood pressure $X_2$  for eleven patients (see Billard and Diday \cite{BillardDiday2000}), Table \ref{Tab1} presents this data set. The objective of the study is to predict $Y$ (response variable) using as predictors $X_1$ y $X_2$. Even if this is not a good example to prove shrinkage methods, since these methods are especially useful when there are many predictor variables, we present this example to compare results with those obtained in Lima Neto and De Carvalho in \cite{Lima2010}. In spite of the above, the results in several indexes are better for shrinkage methods.

\begin{table}[ht] 
\begin{center} 
\begin{tabular}{|c|c|c|c|}
\hline 
\rowcolor[rgb]{0,1,1} & Pulse rate & Systolic blood pressure & Diastolic blood pressure \\ \hline  
\rowcolor[gray]{0.9} 1 &  [44,68] & [90,100] & [50,70] \\ \hline
\rowcolor[gray]{0.9} 2 &  [60,72] & [90,130] & [70,90] \\ \hline
\rowcolor[gray]{0.9} 3 &  [56,90] & [140,180] & [90,100] \\ \hline
\rowcolor[gray]{0.9} 4 &  [70,112] & [110,142] & [80,108] \\ \hline
\rowcolor[gray]{0.9} 5 &  [54,72] & [90,100] & [50,70] \\ \hline
\rowcolor[gray]{0.9} 6 &  [70,100] & [130,160] & [80,110] \\ \hline
\rowcolor[gray]{0.9} 7 &  [63,75] & [140,150] & [60,100] \\ \hline
\rowcolor[gray]{0.9} 8 &  [72,100] & [130,160] & [76,90] \\ \hline
\rowcolor[gray]{0.9} 9 &  [76,98] & [110,190] & [70,110] \\ \hline
\rowcolor[gray]{0.9} 10 &  [86,96] & [138,180] & [90,110] \\ \hline
\rowcolor[gray]{0.9} 11 &  [86,100] & [110,150] & [78,100] \\ \hline
\end{tabular}
\end{center}
\caption{Cardiological interval data set.}
\label{Tab1}
\end{table}

To simplify the data sets tables, we denote {\tt CM} = Center Method, {\tt LassoCM} = Lasso Center Method, {\tt RidgeCM} = Ridge Center Method, {\tt CRM} = Center and Range Method, {\tt LassoCRM} = Lasso Center and Range Method and {\tt RidgeCRM} = Ridge Center and Range Method. Table \ref{Tab2} shows the observed values and fitted values for the variable Pulse Rate using the variables Systolic blood pressure and Diastolic blood pressure as predictors for models {\tt CM}, {\tt LassoCM}, {\tt RidgeCM}, {\tt CRM}, {\tt LassoCRM} y {\tt RidgeCRM}. The values of $\lambda$ for methods {\tt LassoCM}, {\tt RidgeCM}, {\tt LassoCRM} and {\tt RidgeCRM} were estimated using cross-validation; therefore, it is important to clarify that if these methods are executed several times, the results are not necessarily the same. Chosen values were $\lambda = 0.0435635$ for {\tt LassoCM}, $\lambda =  0.875906$ for {\tt LassoCRM}, $\lambda = 0.8752922$ for {\tt RidgeCM} and $\lambda = 875.906$ for {\tt RidgeCRM}.

\begin{table}[ht] 
\begin{small}
\begin{center} 
\begin{tabular}{|c|c|c|c|c|c|c|c|} \hline 
\rowcolor[rgb]{0,1,1} & Pulse rate &  {\tt CM} & {\tt LassoCM} & {\tt RidgeCM} & {\tt CRM} & {\tt LassoCRM} & {\tt RidgeCRM} \\ \hline  
\rowcolor[gray]{0.9} 1 & [44,68] & [59.3,65.9] & [59.4,66] & [60.5,66.8]  & [49.8,75.5] & [53.7,76] & [56.6,78.9] \\ \hline
\rowcolor[gray]{0.9} 2 & [60,72] & [62.7,79.2] & [62.8,79.2] & [63.7,79.1]  & [60.3,81.6] & [60.7,83.1] & [62.2,84.5]  \\ \hline
\rowcolor[gray]{0.9} 3 & [56,90] & [82.5,97.4] & [82.5,97.2] & [82.2,96.1] &81,98.8] &[77.5,99.8] & [74.8,97.2] \\ \hline
\rowcolor[gray]{0.9} 4 & [70,112] & [70.9,86.2] & [70.9,86.2] & [71.4,85.7]& [65.9,91.2]&[67.3,89.6] &[67.3,89.6]\\ \hline
\rowcolor[gray]{0.9} 5 & [54,72] & [59.3,65.9] & [59.4,66] & [60.5,66.8]& [49.7,75.5]&[53.6,76] &[56.6,78.9]\\ \hline
\rowcolor[gray]{0.9} 6 & [70,100] & [77.5,92.5] & [77.5,92.4] & [77.5,91.5]&[71.9,8.1] & [73,95.4]&[71.5,93.9]\\ \hline
\rowcolor[gray]{0.9} 7 & [63,75] & [64.7,79.5] & [64.7,79.5] & [65.6,79.4]&[63.2,81] &[60.2,82.6] &[62.9,85.3]\\ \hline
\rowcolor[gray]{0.9} 8 & [72,100] & [76.8,89.1] & [76.8,89] & [76.9,88.4]& [72.6,93.3]& [71.5,93.8]&[70.2,92.5]\\ \hline
\rowcolor[gray]{0.9} 9 & [76,98] & [69.2,102.3] & [69.3,102.2] & [69.8,100.7]&[74.6,96.9] &[73.9,96.2] &[72.1,94.4]\\ \hline
\rowcolor[gray]{0.9} 10 & [86,96] & [81.8,99.1] & [81.8,98.9] & [81.6,97.7]&[79.9,100.9] &[77.8,100.2] &[75.2,97.6]\\ \hline
\rowcolor[gray]{0.9} 11 & [86,100] & [70.6,87.5] & [70.6,87.4] & [71.1,86.9]&[68,90] &[67.8,90.2] &[67.6,89.9] \\ \hline
\end{tabular}
\end{center}
\caption{Observed and fitted values of Pulse Rate variable by method.}
\label{Tab2}
\end{small}
\end{table}

Table \ref{Tab3} compares the results for {\tt CM}, {\tt LassoCM}, {\tt RidgeCM}, {\tt CRM}, {\tt LassoCRM} and {\tt RidgeCRM}  in indexes the lower boundary root-mean-square error, the upper boundary root-mean-square error square of the lower boundary, the correlation coefficient and the square of the upper boundary correlation coefficient. It can be observed that method {\tt LassoCRM} presents minor RMSE$_L$, the second minor RMSE$_U$ and the second majors $r^2_L$ and $r^2_L$; so in general shrinkage methods tend to reduce the upper and lower root-mean-square errors; and to increase correlations, it can thus be concluded that the methods {\tt LassoCRM} y {\tt RidgeCRM}  are superior in this data set, 
even though it is a table with only two predictor variables.

\begin{table}[ht] 
\begin{center} 
\begin{tabular}{|c|c|c|c|c|}
  \hline
\rowcolor[rgb]{0,1,1}  Method & RMSE$_L$ & RMSE$_U$ & $r^2_L$ & $r^2_U$ \\    \hline  
\rowcolor[gray]{0.9}  {\tt CM} &  11.0942 & 10.41365 & 0.3029147 & 0.5346571  \\   \hline
\rowcolor[gray]{0.9}  {\tt LassoCM} &  11.10846 & 10.42044 & 0.3025499 & 0.534831  \\  \hline
\rowcolor[gray]{0.9}  {\tt RidgeCM} &  11.22309 & 10.52742 & 0.3028591 & 0.534684  \\  \hline 
\rowcolor[gray]{0.9}  {\tt CRM} &  9.809645 & 8.94141 & 0.4153546 & 0.6334484  \\  \hline
\rowcolor[gray]{0.5}  {\tt LassoCRM} &  9.448862 & 9.6991 & 0.4324345 & 0.5583867  \\  \hline
\rowcolor[gray]{0.9}  {\tt RidgeCRM} &  9.584226 & 10.4129 & 0.441123 & 0.5583117  \\  \hline
\end{tabular}
\end{center}
\caption{Performance of the methods on the cardiological data set.}
\label{Tab3}
\end{table}

\subsection{Prostate interval data set}

The original prostate data set come from a study carried out by Stamey et al \cite{Stamey1989} who examined the correlation between the Prostatic Specific Antigen (PSA) level and a series of clinical measures in 97 men who were about to suffer a radical prostatectomy. The objective is to predict the registration of PSA (lpsa) beginning with a series of measures, such as cancer volume (Icavol), the prostate’s weight (lweight), age, the amount of benign hyperplasia of the prostrate (lbph), among others. This is a supervised learning problem, which, since all variables are quantitative it is ideal for the use of linear regression techniques, see \cite{HastieTib2008}.

In order to compare the methods proposed with other regression methods for interval-valued variables and in order to measure the stability of them, we have at random introduced small variations that follow a normal distribution in this data set, with which this data set became an interval-type data set. Also, in order to measure the quality of the predictions, avoiding overfitting, we have divided the data set in two, a training data set size $67 \times 9$, and a test data set size $30 \times 9$ where we will evaluate the models. In Tables \ref{Tab4} and \ref{Tab5} a small portion of both data sets.

\begin{table}[ht] 
\begin{center} 
\begin{tabular}{|c|c|c|c|c|}
  \hline
\rowcolor[rgb]{0,1,1}  & lcavol & lweight & age & $\cdots$ \\    \hline  
\rowcolor[gray]{0.9}  1 &  [-0.5837571,-0.5794132] & [2.766321,2.784411] & [49.99674,50.01089] & $\cdots$  \\   \hline
\rowcolor[gray]{0.9}  2 &  [-0.9955301,-0.9938765] & [3.313408,3.322860] & [57.99987,58.00158] & $\cdots$  \\  \hline
\rowcolor[gray]{0.9}  3 &  [-0.5183475,-0.4818193] & [2.688845,2.695618] & [73.99730,74.00191] &$\cdots$  \\  \hline 
\rowcolor[gray]{0.9}  4 &  [-1.2053568,-1.1674237] & [3.275201,3.289745] & [57.99955,58.00059] & $\cdots$  \\  \hline
\rowcolor[gray]{0.9}  5 &  [0.7341120,0.78214410] & [3.427684,3.441726] & [61.98997,62.00850] &$\cdots$ \\  \hline
\rowcolor[gray]{0.9}  $\vdots$ &  $\vdots$ & $\vdots$ & $\vdots$ & $\ddots$  \\  \hline
\end{tabular}
\end{center}
\caption{Training prostate interval data set.}
\label{Tab4}
\end{table}

\begin{table}[ht] 
\begin{center} 
\begin{tabular}{|c|c|c|c|c|}
  \hline
\rowcolor[rgb]{0,1,1}  & lcavol & lweight & age & $\cdots$ \\    \hline  
\rowcolor[gray]{0.9}  1 &  [0.7312958,0.7626853] & [3.471837,3.482691] & [63.99450,64.00331] & $\cdots$  \\   \hline
\rowcolor[gray]{0.9}  2 &  [-0.7806923,-0.774847] & [3.539259,3.546135] & [57.99987,58.00158] & $\cdots$  \\  \hline
\rowcolor[gray]{0.9}  4 &  [0.2077010,0.2246837] & [3.242688,3.257101] & [62.99144,63.00309] & $\cdots$  \\  \hline
\rowcolor[gray]{0.9}  5 &  [1.2006121,1.2311738] & [3.437803,3.448190] & [56.99564,57.00564] &$\cdots$ \\  \hline
\rowcolor[gray]{0.9}  3 &  [2.0488016 ,2.0636746] & [3.492893,3.509170] & [59.99039,60.00556] &$\cdots$  \\  \hline 
\rowcolor[gray]{0.9}  $\vdots$ &  $\vdots$ & $\vdots$ & $\vdots$ & $\ddots$  \\  \hline
\end{tabular}
\end{center}
\caption{Testing prostate interval data set.}
\label{Tab5}
\end{table}

Using the data of Table \ref{Tab4} linear regression models were built with {\tt CM}, {\tt LassoCM}, {\tt RidgeCM}, {\tt CRM}, {\tt LassoCRM} y {\tt RidgeCRM} methods. Then, these methods were evaluated with the testing data set of Table \ref{Tab5}. Table \ref{Tab6} compares the results for the indexes: the lower boundary root-mean-square error, the upper boundary root-mean-square error square, the square of the lower boundary of the correlation coefficient and the square of the upper boundary correlation coefficient.

\begin{table}[ht] 
\begin{center} 
\begin{tabular}{|c|c|c|c|c|}
  \hline
 \rowcolor[rgb]{0,1,1} Method & RMSE$_L$ & RMSE$_U$ & $r^2_L$ & $r^2_U$ \\    \hline  \hline
 \rowcolor[gray]{0.9} CM &  0.7229999 & 0.7192467 & 0.501419 & 0.5058389 \\   \hline
 \rowcolor[gray]{0.5} LassoCM &  0.6945169 & 0.6914335 & 0.5409863 & 0.544571  \\  \hline
 \rowcolor[gray]{0.9}  RidgeCM &  0.703543 & 0.7004145 & 0.5286114& 0.5322683  \\  \hline \hline
 \rowcolor[gray]{0.9} CRM &  0.7212187 & 0.7209186 & 0.5034327 & 0.5039147  \\  \hline
 \rowcolor[gray]{0.5} LassoCRM &  0.6995754 & 0.6999873 & 0.5332342 & 0.5327807  \\  \hline
\rowcolor[gray]{0.5}  RidgeCRM &  0.698668 & 0.6989728 & 0.5354419& 0.5351355 \\  \hline
\end{tabular}
\end{center}
\caption{Performance of the methods in prostate test interval data set.}
\label{Tab6}
\end{table}

\begin{figure}[htbp]
\begin{center}
\includegraphics[width=9cm]{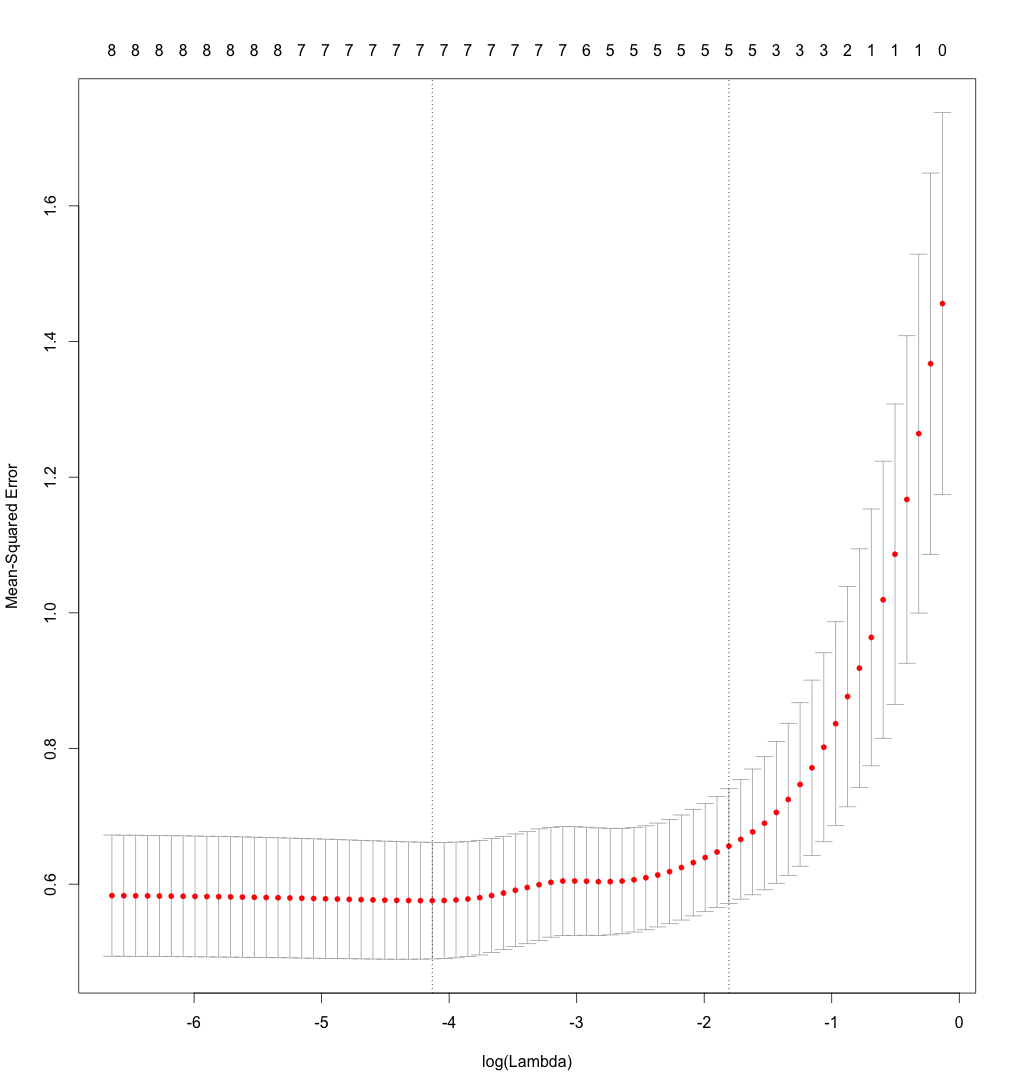} 
\caption{Choice of $\lambda$ using cross-validation. \label{Fig1}}
\end{center}
\end{figure}

In Table \ref{Tab6} we can observe that  {\tt LassoCM}, {\tt RidgeCRM} and  {\tt LassoCRM} methods present minor RMSE$_L$ and RMSE$_U$. It is also observed that the {\tt LassoCM} method presents the major $r^2_L$ and $r^2_L$ and {\tt LassoCRM} method presents the third major $r^2_L$ and $r^2_L$, followed by  {\tt RidgeCM}, so shrinkage methods tend to reduce quadratic errors RMSE$_L$ and RMSE$_U$; and to increase correlations $r^2_L$ and $r^2_L$ so {\tt LassoCM}, {\tt LassoCRM} and {\tt RidgeCRM} methods are superior in this data set. As observed in Figure \ref{Fig1}, the above results for the method {\tt LassoCM} were obtained using a cross-validation process to obtain the best possible value for $\lambda$ from where $\ln(\lambda)=$1.75 was taken, that is $\lambda=$5.7546.

\begin{figure}[htbp]
\begin{center}
\includegraphics[width=9cm]{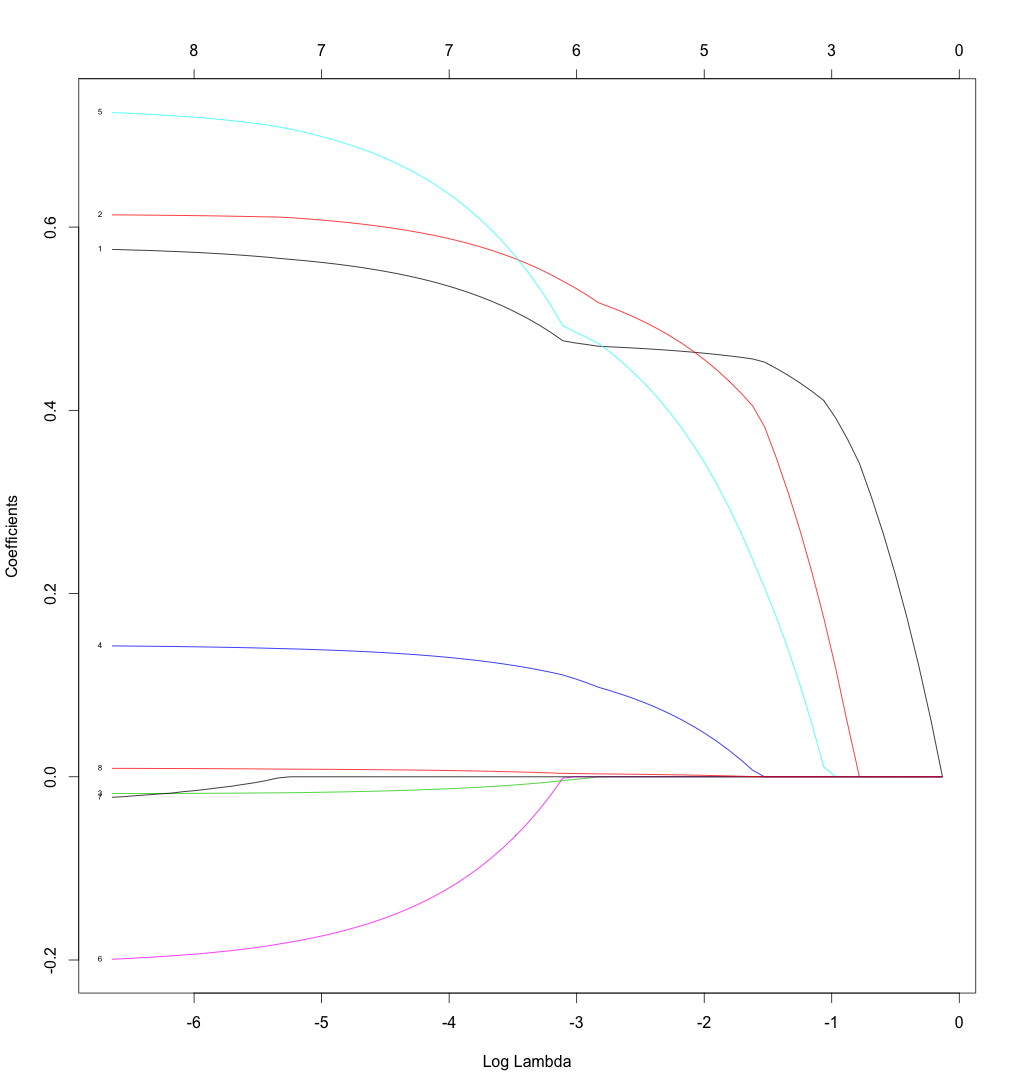} 
\caption{{\tt LassoCM} variables selection. \label{Fig2}}
\end{center}
\end{figure}

For this $\lambda$ value, as observed in Figure \ref{Fig2},  {\tt LassoCM} method have selected four predictors, predictor 5 (svi), predictor 2 (lweight), predictor 1 (Icavol) and predictor 4 (lbph), therefore, the regress equation fitted by the {\tt LassoCM} method for the training prostate interval data set is:

\[ \hat{Y}_{\tt LassoCM}(\text{lpsa})  \approx \text{0.56} +  \text{0.58} \cdot \text{lcavol}+\text{0.62} \cdot \text{lweight}+\text{0.168} \cdot \text{lbph}+\text{0.667} \cdot \text{svi}, \]

\noindent while the fitted regression equation fitted by the CM method for the training prostate interval data set is: 

\begin{eqnarray*}
\hat{Y}_{\tt CM}(\text{lpsa}) &\approx& \text{0.4}+ \text{0.6} \cdot \text{lcavol} +  \text{0.6} \cdot \text{lweight}   - \text{0.02}  \cdot \text{age}  + \text{0.1} \cdot \text{lbph} + \\
 &&  \text{0.7} \cdot \text{svi} - \text{0.2} \cdot \text{lcp}  - \text{0.03} \cdot \text{gleason} + \text{0.009} \cdot  \text{pgg45}.
\end{eqnarray*}
                           
\noindent It being clear that the {\tt LassoCM} method, besides  producing better results, provides a regression equation much easier to interpret, since it only uses four variables.

Using training prostate interval data set shown in Table \ref{Tab4}, linear regression models with the elastic net center method (NetCM) and elastic net center and range method (NetCRM) were fitted. The best results were obtained taking $\alpha=$0.9 and $\alpha=$0.8 respectively, by varying $\alpha \in (\text{0,0.1,0.2,0.3,0.4,0.5,0.6,0.7,0.8,0.9,1})$. Results obtained are shown in Table \ref{Tab7}, the quality of the prediction is quite similar to that achieved with {\tt LassoCM} and {\tt LassoCRM} methods.

\begin{table}[ht] 
\begin{center} 
\begin{tabular}{|c|c|c|c|c|}
  \hline
 \rowcolor[rgb]{0,1,1} Method & RMSE$_L$ & RMSE$_U$ & $r^2_L$ & $r^2_U$ \\    \hline  \hline
 \rowcolor[gray]{0.9} NetCM &  0.6996586 & 0.6964444 & 0.5336516 & 0.53742 \\   \hline
 \rowcolor[gray]{0.9} NetCRM &  0.6989958 & 0.699399 & 0.534105 & 0.5336628  \\  \hline
\end{tabular}
\end{center}
\caption{Performance of elastic shrinkage net methods in prostate test interval data set.}
\label{Tab7}
\end{table} 

\subsection{US Murder interval data set}

These data were taken from UCI Machine Learning Repository, consult \cite{Bache2013} and  \cite{Rod2014}. The idea is to study this murders table in $n = 1994$ communities of the United State, for this we have $p = 103$ variables measured over each one of these communities. The objective is to predict the mortality rate (ViolentCrimesPerPop) as a linear function of these $p = 103$ predictors.

The first action taken to analyze this data set was to convert it in a symbolic data set where all predictors are of the interval type, taking for this as concept the variable {\tt State}, that is, each state of US will be a statistic unit that we will study. We will thus have 46 rows (states present in US Murder data set) and 102 interval-valued predictors. This is a data set rather adequate to measure the affectivity of shrinkage center and range methods since there is an important amount of predictors. Table \ref{Tab8} presents partially the US Murder classic data set, this table being $1994 \times 103$ and Table \ref{Tab9} presents partially the US Murder interval data set, the size of this table is $46 \times 102$. Chapter 5 of the book \cite{BockDiday2000}  explains in detail how to transform a classic data table in a symbolic data table. 

\begin{table}[ht] 
\begin{footnotesize}
\begin{center} 
\begin{tabular}{|c|c|c|c|c|c|c|c|c|}
  \hline
\rowcolor[rgb]{0,1,1}  N & state & fold & population & householdsize & racepctblack & racePctWhite & racePctAsian & $\cdots$ \\    \hline  
\rowcolor[gray]{0.9}  1 &  8  &  1 &      0.19 &         0.33   &      0.02    &     0.90     &    0.12 & $\cdots$  \\   \hline
\rowcolor[gray]{0.9}  2  &  53 &   1   &    0.00   &       0.16  &       0.12  &       0.74    &     0.45 & $\cdots$  \\   \hline
\rowcolor[gray]{0.9}  3   & 24  &  1   &    0.00    &      0.42  &       0.49    &     0.56     &    0.17 & $\cdots$  \\   \hline
\rowcolor[gray]{0.9}  4  &  34   & 1   &    0.04    &      0.77  &       1.00    &     0.08      &   0.12 & $\cdots$  \\   \hline
\rowcolor[gray]{0.9}  5   & 42   & 1   &    0.01     &     0.55   &      0.02    &     0.95     &    0.09 & $\cdots$  \\   \hline
\rowcolor[gray]{0.9}  6   &  6  &  1   &    0.02      &    0.28    &     0.06     &    0.54      &   1.00 & $\cdots$  \\   \hline
\rowcolor[gray]{0.9}  7   & 44   & 1  &     0.01      &    0.39    &     0.00    &     0.98      &   0.06 & $\cdots$  \\   \hline
\rowcolor[gray]{0.9}  8   &  6   & 1   &    0.01      &    0.74     &    0.03     &    0.46     &    0.20 & $\cdots$  \\   \hline
\rowcolor[gray]{0.9}  $\vdots$ &  $\vdots$ & $\vdots$ & $\vdots$ & $\vdots$ & $\vdots$ & $\vdots$ & $\vdots$ & $\cdots$  \\  \hline
\rowcolor[gray]{0.9}  1994   &  6  & 10   &    0.2      &    0.78     &    0.14    &    0.46     &    0.24 & $\ddots$  \\   \hline
\end{tabular}
\end{center}
\caption{US Murder classic data set.}
\label{Tab8}
\end{footnotesize}
\end{table}

\begin{figure}[htbp]
\begin{center}
\includegraphics[width=9cm]{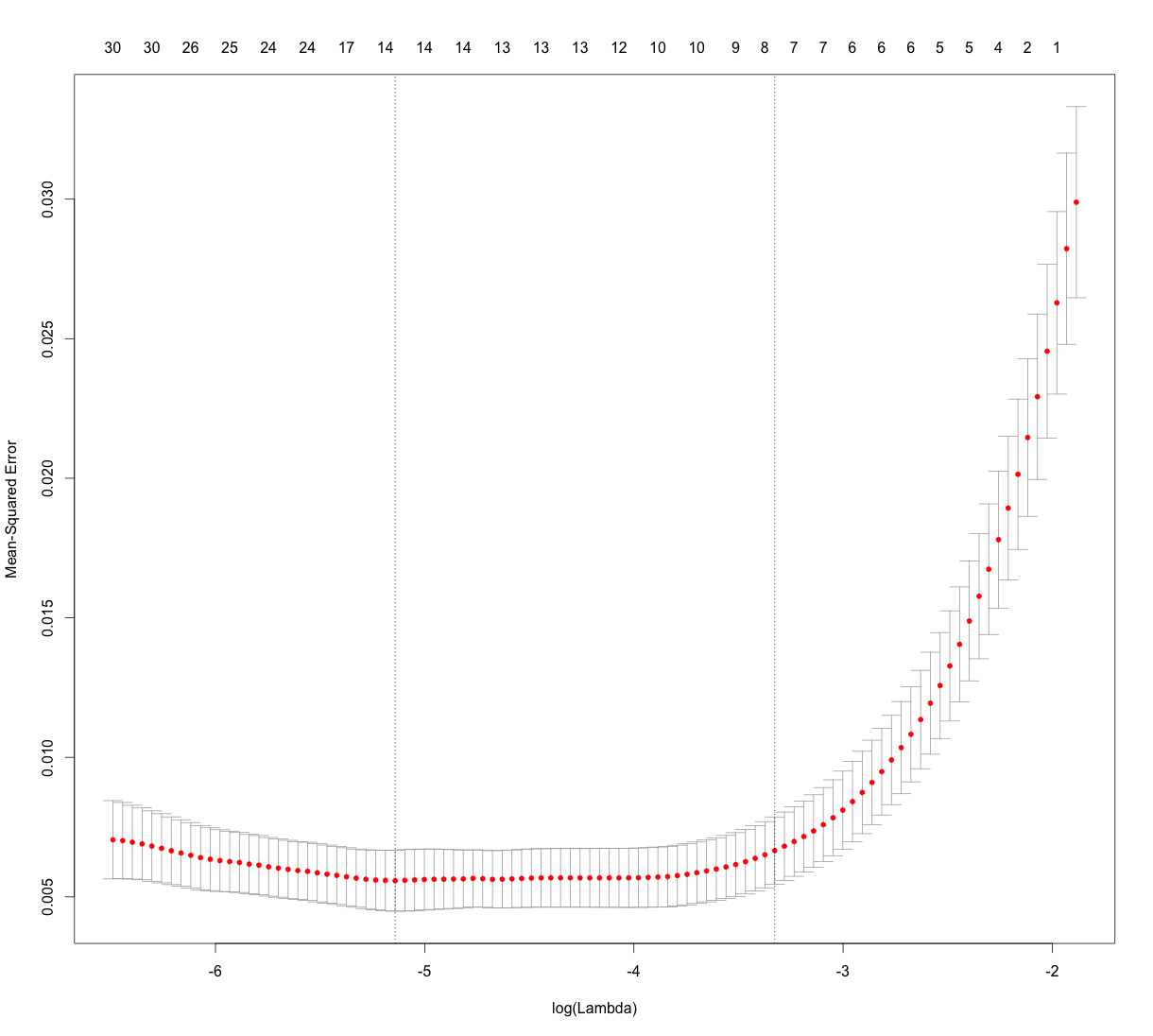} 
\caption{Choice of $\lambda$ using cross-validation. \label{Fig3}}
\end{center}
\end{figure}

\begin{figure}[htbp]
\begin{center}
\includegraphics[width=9cm]{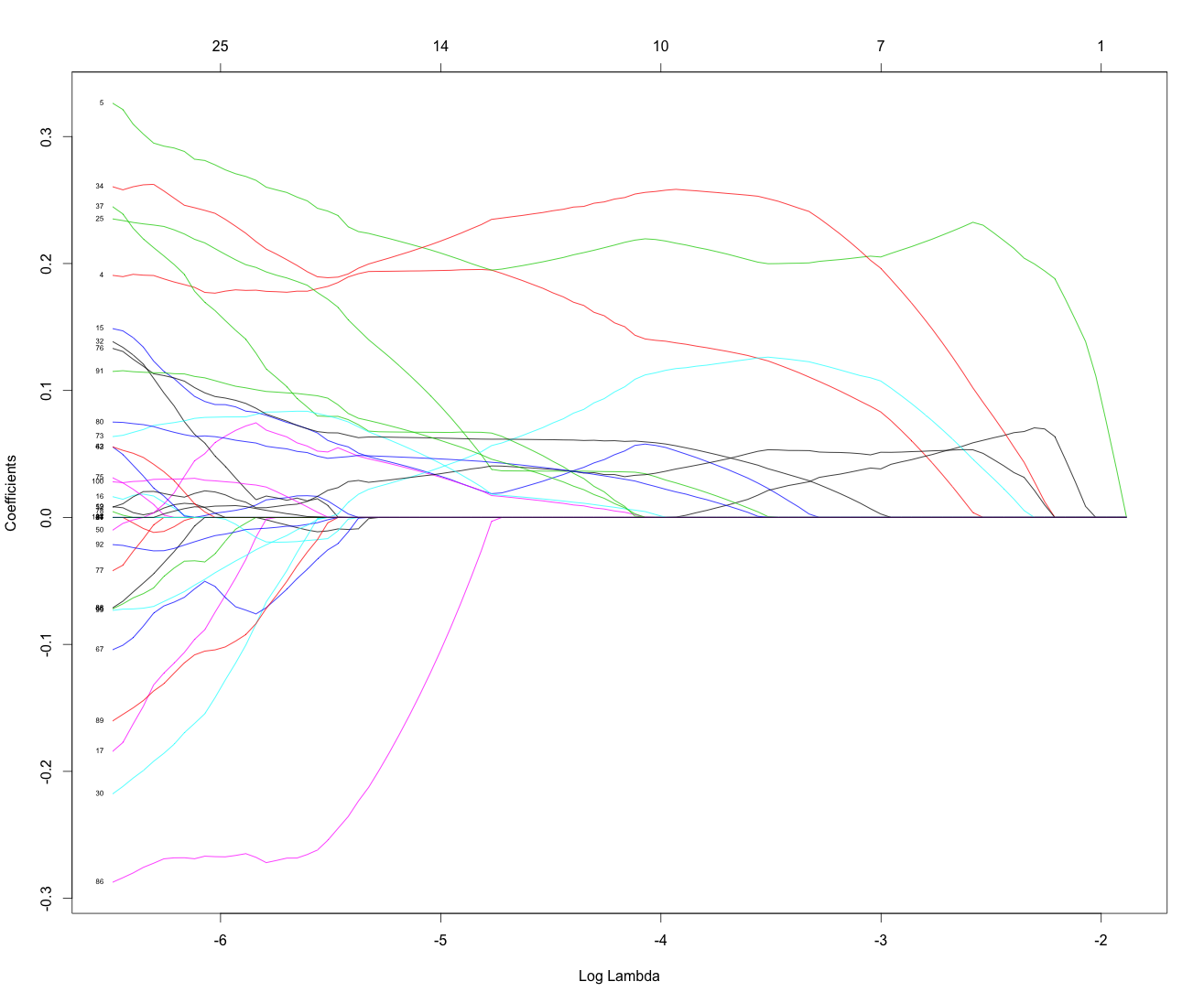} 
\caption{{\tt LassoCRM} variables selection. \label{Fig4}}
\end{center}
\end{figure}

\begin{figure}[htbp]
\begin{center}
\includegraphics[width=9cm]{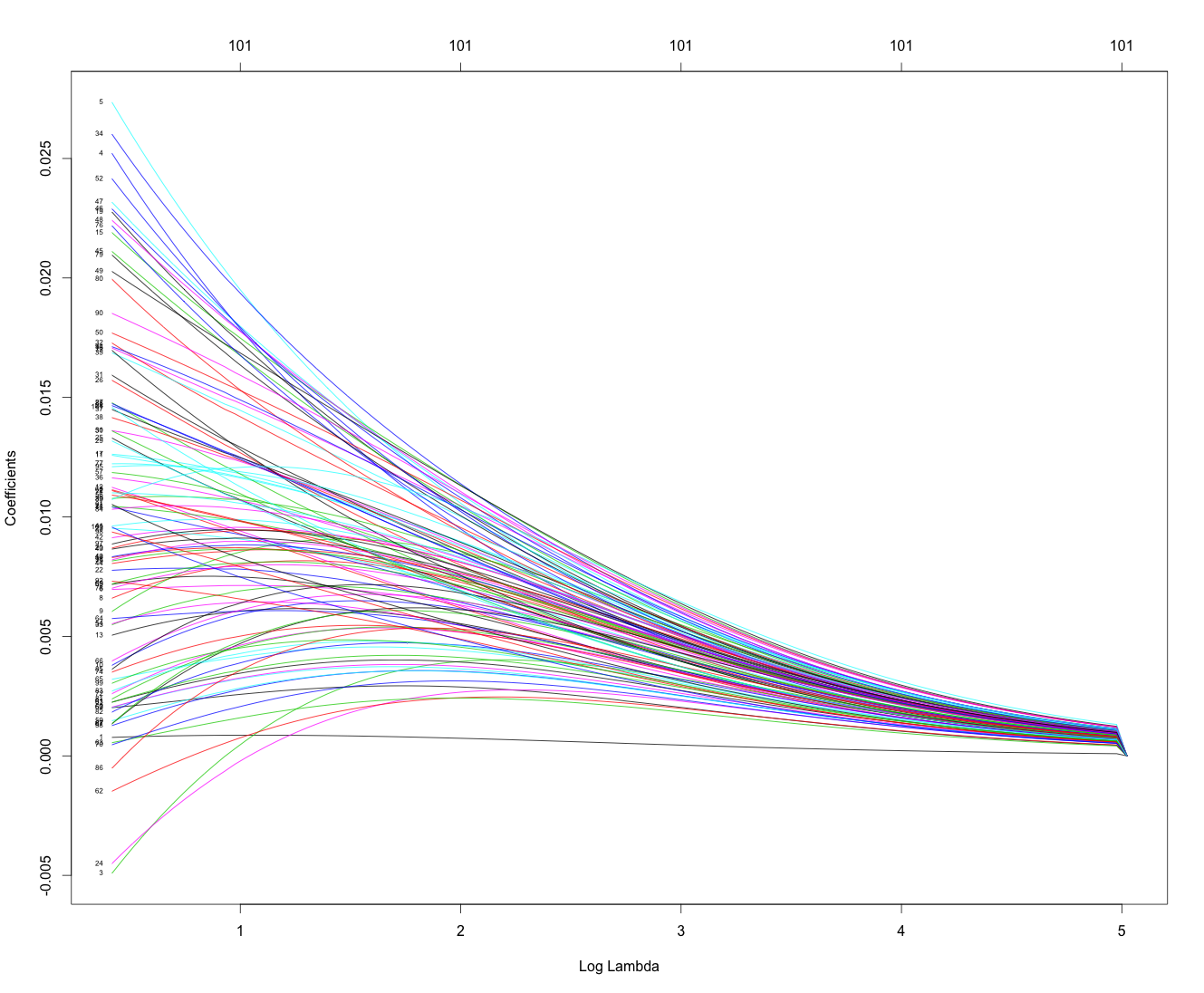} 
\caption{{\tt RidgeCRM} variables contraction coefficients. \label{Fig6}}
\end{center}
\end{figure}

In Table \ref{Tab10} we observe that {\tt LassoCRM} method presents the less errors RMSE$_L$ and RMSE$_U$  and the major correlations $r^2_L$ y $r^2_L$ followed by {\tt RidgeCRM} method, we can then affirm that {\tt LassoCRM} has a significantly better performance than the remaining methods in this data set, followed by {\tt RidgeCRM}. As observed in Figure \ref{Fig3} the above results for {\tt LassoCRM} were obtained using a cross-validation process to obtain the best possible value for $\lambda$, from where, it was taken $\ln(\lambda)=-$4, that is, $\lambda=$0.0183156. For this value of $\lambda$, as observed in Figure \ref{Fig4},  {\tt LassoCRM}  have selected 9 predictors from 102 predictors in the data set, that is, besides {\tt LassoCRM}  is producing results significantly better than the other methods, it generates results much easier to interpret, since it used in the linear fitment equation only 9 predictors instead of 102 predictors as done by {\tt CM}, {\tt CRM}, {\tt RidgeCM} and {\tt RidgeCRM} methods. Figure \ref{Fig6} illustrates the shrinkage process of the coefficients of the regression equation in {\tt RidgeCRM} model.

\begin{table}[ht] 
\begin{small}
\begin{center} 
\begin{tabular}{|c|c|c|c|c|c|c|c|}
  \hline
\rowcolor[rgb]{0,1,1}   state & fold & population & householdsize & racepctblack & racePctWhite & racePctAsian & $\cdots$ \\    \hline  
\rowcolor[gray]{0.9}  1 &  [1,10]   &  [0,0.41] & [0.23,0.67] & [0,1.00] &  [0,0.99]  & [0,0.21] &  $\cdots$  \\   \hline
\rowcolor[gray]{0.9}  2 &  [4,9]   &  [0.03,0.35] & [0.46,0.53] & [0.02,0.25] &  [0.58,0.71]  & [0.2,0.3] &  $\cdots$  \\   \hline
\rowcolor[gray]{0.9}  3 &  [1,10]   &  [0,1] & [0.23,0.98] & [0,0.23] &  [0.37,0.97]  & [0.02,0.32] &  $\cdots$  \\   \hline
\rowcolor[gray]{0.9}  4 &  [1,10]   &  [0,0.27] & [0.22,0.59] & [0,1] &  [0.12,1]  & [0.01,0.25] &  $\cdots$  \\   \hline
\rowcolor[gray]{0.9}  5 &  [1,10]   &  [0,1] & [0,1] & [0,1] &  [0,0.96]  & [0.03,1] &  $\cdots$  \\   \hline
\rowcolor[gray]{0.9}  6 &  [1,9]   &  [0,0.74] & [0.21,0.6] & [0,0.25] &  [0.58,0.95]  & [0.01,0.24] &  $\cdots$  \\   \hline
\rowcolor[gray]{0.9}  $\vdots$ &  $\vdots$ & $\vdots$ & $\vdots$ & $\vdots$ & $\vdots$ & $\vdots$  & $\cdots$  \\  \hline
\rowcolor[gray]{0.9}  46 &  [3,9]   &  [0,0.06] & [0.29,0.67] & [0,0.06] &  [0.85,0.97]  & [0.02,0.14] &  $\ddots$  \\   \hline
\end{tabular}
\end{center}
\caption{US Murder interval data set.}
\label{Tab9}
\end{small}
\end{table}

\begin{table}[ht] 
\begin{center} 
\begin{tabular}{|c|c|c|c|c|}
  \hline
\rowcolor[rgb]{0,1,1}  Method & RMSE$_L$ & RMSE$_U$ & $r^2_L$ & $r^2_U$ \\    \hline  \hline
\rowcolor[gray]{0.9} CM &  0.3177475 & 0.3571175 & 0.8052504 & 0.6235015 \\   \hline
\rowcolor[gray]{0.9}  LassoCM &  0.2203994 & 0.3360708 & 0.8871075& 0.7486212  \\  \hline
\rowcolor[gray]{0.9}  RidgeCM &  0.1948507 & 0.2533318 & 0.7077219 & 0.7909678  \\  \hline 
\rowcolor[gray]{0.9}  CRM &  0.06576857 & 0.1701506 & 0.831561 & 0.685454  \\  \hline
\rowcolor[gray]{0.5}  LassoCRM &  0.03322222 & 0.09115312 & 0.9568711 & 0.915639  \\  \hline
\rowcolor[gray]{0.7}  RidgeCRM &  0.05344275 & 0.1549643 & 0.9074878 & 0.8024979 \\  \hline
\end{tabular}
\end{center}
\caption{Performance of the methods in US Murder interval data set.}
\label{Tab10}
\end{table}

In case of table US Murder interval data set, in Table \ref{Tab11} we observe that the linear regression model fitted with the elastic net center and range method (NetCRM) with  $\alpha=$0.9 slightly improves the results obtained with {\tt LassoCRM}.

\begin{table}[ht] 
\begin{center} 
\begin{tabular}{|c|c|c|c|c|}
  \hline
 \rowcolor[rgb]{0,1,1} Method & RMSE$_L$ & RMSE$_U$ & $r^2_L$ & $r^2_U$ \\    \hline  \hline
 \rowcolor[gray]{0.9} NetCRM &   0.03315683 & 0.09084465  & 0.9570477 & 0.916268  \\  \hline
\end{tabular}
\end{center}
\caption{Performance of elastic shrinkage net methods in US Murder interval data set.}
\label{Tab11}
\end{table} 

\section{Conclusions and future work}
\label{sec:cr}

In this paper we have proposed six new methods to fit linear regression for interval-valued variables: {\tt RidgeCM}, {\tt RidgeCRM}, {\tt LassoCM}, {\tt LassoCRM}, {\tt NetCM} and {\tt NetCRM}, all based on the central idea of fit the regression equations for the centers and for the ranges of intervals applying shrinkage linear methods. In the case {\tt LassoCM} and {\tt LassoCRM} this generalization is more complicated since the fitment of the regression model over the ranges should be restricted to the predictors selected by the Lasso method in the fitted regression model over the intervals' midpoints.

The experimental analysis presented in section  \ref{sec:ee} permitted us to verify in three real data sets, cardiological interval data set, prostrate interval data set and US murder interval data set, that the use of shrinkage linear methods improve in an important way the results of the predictions, especially in US murder interval data set since this data set has a much larger quantity of predictors. It has also been verified that linear generation regression equations generated with {\tt LassoCRM} have also the treat advantage of producing solutions easier to interpret thanks to the automatic variables selection carried out.

The methods proposed in this work, just like {\tt CM} and {\tt CRM} methods, have the problem explained by Lima Neto and De Carvalho in \cite{Lima2010}, that it cannot be guarantee that mathematically $\hat{y}_{Li} \le \hat{y}_{Ui}$ for all $i=1,\ldots n$. In future works we will be applying the idea proposed in \cite{Lima2010}, which consists in generating the linear regression models using certain restrictions that will permit us to guarantee that proposed methods in this work satisfy that $\hat{y}_{Li} \le \hat{y}_{Ui}$. This was not included in this research since it could cause confusion in the results, since it could not be clear if improvements in predictions are due to the applications of shrinkage or rather to the application of restrictions in the regression methods.

\end{document}